\documentclass[
aps,%
12pt,%
final,%
notitlepage,%
oneside,%
onecolumn,%
nobibnotes,%
nofootinbib,% In the current version of REVTeX, when this option is on,
superscriptaddress,%
noshowpacs,%
centertags]%
{revtex4}

\usepackage{graphicx}
\usepackage{subfigure}
\usepackage{sidecap}

\begin{document}
\title{Core collapse supernovae in the QCD phase diagram}

\author{\firstname{T.}~\surname{Fischer}}
\email{t.fischer@gsi.de}
\affiliation{GSI, Helmholtzzentrum f\"ur Schwerionenforschung GmbH,
Darmstadt, Germany}
\affiliation{Technische Universit{\"a}t Darmstadt, Germany}

\author{\firstname{D.}~\surname{Blaschke}}
\affiliation{Institute for Theoretical Physics, University of Wroclaw, Poland}
\affiliation{Bogoliubov  Laboratory of Theoretical Physics, JINR Dubna, Russia}

\author{\firstname{M.}~\surname{Hempel}}
\affiliation{Department of Physics, University of Basel, Switzerland}

\author{\firstname{T.}~\surname{Kl\"ahn}}
\affiliation{Institute for Theoretical Physics, University of Wroclaw, Poland}

\author{\firstname{R.}~\surname{\L astowiecki}}
\affiliation{Institute for Theoretical Physics, University of Wroclaw, Poland}

\author{\firstname{M.}~\surname{Liebend\"orfer}}
\affiliation{Department of Physics, University of Basel, Switzerland}

\author{\firstname{G.}~\surname{Mart{\'i}nez-Pinedo}}
\affiliation{GSI, Helmholtzzentrum f\"ur Schwerionenforschung GmbH,
Darmstadt, Germany}
\affiliation{Technische Universit{\"a}t Darmstadt, Germany}

\author{\firstname{G.}~\surname{Pagliara}}
\affiliation{Institut f\"ur Theoretische Physik,
Ruprecht-Karls-Universit\"at, Heidelberg, Germany}

\author{\firstname{I.}~\surname{Sagert}}
\affiliation{Institut f\"ur Theoretische Physik,
Ruprecht-Karls-Universit\"at, Heidelberg, Germany}

\author{\firstname{F.}~\surname{Sandin}}
\affiliation{Department of Computer Science and Electrical Engineering,
EISLAB, Lule{\aa} Tekniska Universitet, Lule{\aa}, Sweden}
\affiliation{D\'{e}partement AGO-IFPA, Universit\'{e} Liege, Belgium}

\author{\firstname{J.}~\surname{Schaffner-Bielich}}
\affiliation{Institut f\"ur Theoretische Physik,
Ruprecht-Karls-Universit\"at, Heidelberg, Germany}

\author{\firstname{S.}~\surname{Typel}}
\affiliation{GSI, Helmholtzzentrum f\"ur Schwerionenforschung GmbH,
Darmstadt, Germany}
\affiliation{Excellence Cluster Universe,
Technische Universit\"{a}t M\"{u}nchen, Germany}

\begin{abstract}
We compare two classes of
hybrid equations of state with a hadron-to-quark matter 
phase transition in their application to core collapse 
supernova simulations.
The first one uses the quark bag model and describes the
transition to three-flavor quark matter at low critical densities.
The second one employs a Polyakov-loop extended Nambu--Jona--Lasinio
(PNJL) model with parameters describing a phase transition to
two-flavor quark matter at higher critical densities.
These models possess a distinctly different temperature dependence of 
their transition densities which turns out to be crucial for the possible 
appearance of quark matter in supernova cores. 
During the early post bounce accretion phase quark matter is found only
if the phase transition takes place at sufficiently low densities as in the 
study based on the bag model.
The increase critical density with increasing temperature,
as obtained for our PNJL parametrization,
prevents the formation of quark  matter.
The further evolution of the core collapse supernova as obtained
applying the quark bag model leads to a structural reconfiguration
of the central  protoneutron star where,
in addition to a massive pure quark matter core,
a strong hydrodynamic shock wave forms and a second neutrino
burst is released during the shock propagation across the neutrinospheres.
We discuss the severe constraints in the freedom of choice of quark matter 
models and their parametrization due to the recently observed 2~M$_\odot$ 
pulsar and their implications for further studies of core collapse supernovae
in the QCD phase diagram.
\end{abstract}

\maketitle

\section{Introduction}

Stars more massive than 8~M$_\odot$ explode as core collapse
supernovae, with kinetic explosion energies of the ejected material
on the order of $10^{51}$~erg.
The remnants, the protoneutron stars (PNSs), are initially hot and
lepton-rich and cool via deleptonization during the first 30~seconds
after the onset of the explosion.
Above a certain progenitor mass threshold on the order of 40~M$_\odot$,
which is an active subject of research, stars will no longer explode.
Such models will collapse and from black holes.
The critical mass for a PNS to collapse and from a black hole
is given by the equation of state (EoS).
The commonly used EoS in core collapse supernova simulations
are based on pure hadronic descriptions,
e.g. the compressible liquid-drop model with surface effects
\cite{Lattimer:1991nc}
and relativistic mean field theory including the Thomas-Fermi
approximation for heavy nuclei \cite{Shen:1998by}.
The conditions that are obtained in PNS interiors during the simulation,
are several times nuclear matter density,
temperatures on the order of several tens of MeV and a low
proton-to-baryon ratio given by the electron fraction
of\footnote{The proton-to-baryon ratio, $Y_p=n_p/n_B$,
is equal to the electron fraction, 
$Y_e:=Y_{e^-}-Y_{e^+} $,
in the absence of muons.} $Y_e=Y_p\leq0.3$.
Under such conditions, the quark-hadron phase transition is not unlikely to
take place and the assumption of pure hadronic matter becomes
questionable.
Ab initio calculations of the phase diagram and EoS of quantum 
chromodynamics (QCD) as the fundamental theory of strongly interacting 
matter come from simulations of this gauge theory on the lattice but are 
still restricted to low baryon densities (chemical potentials). 
In this region of the phase diagram they predict a crossover transition 
with a pseudocritical temperature for chiral symmetry restoration and 
deconfinement at $T_c\simeq 150 - 170$ MeV 
\cite{Borsanyi:2010bp,Bazavov:2010bx}.
A critical endpoint for first order transitions is conjectured but lies,
if it exists at all, outside the region presently accessible by lattice QCD.
An interesting conjecture supported by a statistical model analysis of 
hadron production in heavy-ion collisions and by the large-N$_c$ limit of 
QCD suggests the existence of a triple point in the phase diagram
\cite{Andronic:2009gj} due to a third, ``quarkyonic'' phase at temperatures
$T<T_c$ and high baryon densities \cite{McLerran:2007qj}. 
This state of matter might become accessible in experiments at the planned 
third generation of heavy-ion collision facilities FAIR in Darmstadt
(Germany) and NICA in Dubna (Russia) \cite{Blaschke:2010ka} 
which thus allow systematic laboratory studies of conditions like in supernova 
collapse and protoneutron star evolution  \cite{Klahn:2011au}.

In order to investigate the appearance of quark matter in core collapse
supernova models, the implementation of a quark-hadron hybrid EoS is required.
It must be valid for a large range of densities $n_B$, temperatures $T$
and proton-to-baryon ratios.
At large baryon densities, where lattice QCD cannot be applied due
to current conceptional limitations, phenomenological models are commonly used.
In the present study we will discuss hybrid EoS which employ 
quark matter models that are representative examples from
two wide classes: bag models and NJL-type models.
The popular and simple thermodynamical bag model is inspired by the success
of the vacuum MIT bag model for the hadron spectrum \cite{DeTar:1983rw}.
It describes quarks as non-interacting fermions of a constant mass confined 
by an external ``bag'' pressure $B$.
NJL-type models are constructed to obey basic symmetries of the QCD Lagrangian 
like the chiral symmetry of light quarks, and to describe its dynamical 
breaking which results in medium-dependent masses (see \cite{Buballa:2003qv} 
and references therein). 
The inclusion of diquark interaction channels leads to a rich phase structure 
at low temperatures and high densities with color superconductivity (diquark
condensation) in two-flavor (2SC) and three-flavor (CFL) quark matter
(see, e.g., \cite{Blaschke:2005uj} for the phase structure and 
\cite{Sandin:2007zr} for the relevance to protoneutron stars;
detailed information concerning color superconductivity is
found in review articles \cite{Buballa:2003qv,Alford:2007xm}). 
These models have no confining interaction and would therefore lead to the 
unphysical dominance of thermal quark excitations already at temperatures 
well below $T_c$. Their coupling to the Polyakov loop potential is essential 
to suppress the unphysical degrees of freedom and it can be adjusted
to fit the behavior of lattice QCD thermodynamics at low densities
\cite{Roessner:2006xn}. Extending this effective model to high densities 
leads to the class of PNJL models of which we apply one for this study.

Different assumptions made for the description of quark matter
lead to different critical conditions for the onset of deconfinement,
which are given in terms of a critical density that depends on the
temperature and the proton-to-baryon ratio.
The resulting different phase diagrams may lead to a different
evolutionary behavior for core collapse supernovae.
The central supernova conditions start from low densities
($6\times10^{-6}$~fm$^{-3}$/$10^{10}$~g cm$^{-3}$)
and temperatures (0.6~MeV),
where pure hadronic matter dominates,
and approach densities above nuclear saturation density
(0.16~fm$^{-1}$/$2.7\times10^{14}$~g cm$^{-3}$)
at temperatures on the order to tens of MeV and $Y_p\simeq0.2-0.3$.
The nature of the QCD transition is not precisely understood on a microscopic
level. Therefore, one usually splices independent nuclear and quark matter 
EoS by constructing the phase transition using more or less appropriate 
conditions for the phase equilibrium. 
Popular examples are the Maxwell, Gibbs or Glendenning  
\cite{Glendenning:1992vb} constructions, whereby usually the Maxwell 
construction leads to the smallest region between critical densities for the 
onset and the end of the mixed phase.
We will disregard finite size effects (pasta structures) and also 
non-equilibrium effects due to the nucleation of the new phase with the 
justification that weak processes establishing the chemical equilibrium
are fast compared to the typical timescales encountered in supernova
simulations \cite{Mintz:2009ay}.
We will compare the evolution of a representative core collapse supernova
simulation in the two different phase diagrams based on the bag model
and the PNJL model.
The core collapse supernova model is based on general relativistic
radiation hydrodynamics and three flavor Boltzmann neutrino
transport
\cite{Mezzacappa:1993gm,Mezzacappa:1993gn,Mezzacappa:1993,Liebendoerfer:2000cq,Liebendoerfer:2000fw,Liebendoerfer:2002xn}

The manuscript is organized as follows: In Sect. 2 we introduce the standard
core collapse supernova phenomenology.
In Sect. 3 we briefly introduce the two quark-hadron hybrid EoS,
the quark bag model and the PNJL model.
The evolution of a representative 15~M$_\odot$ core collapse supernova
model in the phase diagram is illustrated in Sect. 4 by comparing the quark bag
and PNJL models.
We close with a summary in Sect. 5.

\section{The standard scenario of core collapse supernovae}

Si-burning produces Fe-cores at the final phase of stellar evolution
of massive stars.
These Fe-cores start to contract due to the photodisintegration of heavy
elements and electron captures.
The latter reduces the dominant pressure of the degenerate electron gas.
During the collapse, density and temperature rise and hence
electron captures, which deleptonize the central core, increase.
The collapse accelerates until neutrino trapping densities,
on the order of $\rho\simeq10^{11}$--$10^{13}$~g/cm$^3$,
are obtained after which the collapse proceeds adiabatically.
At nuclear matter density, the repulsive nuclear interaction stiffens
the EoS significantly and the collapse halts.
The core bounces back, where a sound wave forms,
which steepens into a shock wave.
The shock wave propagates outwards where the dissociation of infalling
heavy nuclei causes an energy loss of about 8~MeV per baryon.
Furthermore, during the shock propagation across the neutrinospheres,
which are the neutrino energy and flavor dependent spheres of last scattering,
additional electron captures release a burst of $\nu_e$
that carries away 4--$5\times10^{53}$~erg/s
on a short timescale of 5--20~ms post bounce
(depending on the progenitor model and the EoS).
Both sources of energy loss turn the dynamic shock into a
standing accretion shock, already at about 5~ms post bounce.

The post bounce evolution is determined by mass accretion, due to the
continuously infalling material from the outer layers of the Fe-core as
well as the surrounding Si-layer (depending on the progenitor model).
On a timescale on the order of 100~ms up to seconds,
the central density and temperature increase continuously.
In order to achieve an explosion, energy needs to be deposit
behind the standing accretion shock which subsequently
revives the standing accretion shock.
Several mechanisms have been suggested, including
the magnetically-driven
\cite{LeBlanc:1970kg,Moiseenko:2007,Takiwaki:2009},
the acoustic \cite{Burrows:2006}
and the neutrino-driven \cite{Bethe:1984ux}.
The standard scenario, delayed explosions due to neutrino heating,
have been shown to work in spherical symmetry
\cite{Kitaura:2005bt,Fischer:2009af}
for the low mass 8.8~M$_\odot$ ONeMg-core
\cite{Nomoto:1983,Nomoto:1984,Nomoto:1987}.
For more massive progenitors, multi-dimensional phenomena such as
rotation and the development of fluid instabilities are required  and
help to increase the neutrino heating efficiency
\cite{Miller:1993,Herant:1994dd,Burrows:1995ww,Janka:1996}.
Such models are also required to aid the understanding
of aspherical explosions
\cite{Bruenn:2009,Marek:2009}.

\section{The quark bag and PNJL hybrid models}

We discuss two different quark matter descriptions for use in astrophysical
applications as well as their differences in the resulting EoS.
The quark bag model for three flavor quark matter is described in detail
in the  Refs.~\cite{Sagert:2008ka, Fischer:2010zzb, Fischer:2010wp}.
Bag constants in the range of $B^{1/4}=$155--165~MeV and
corrections from the strong interaction were adopted
and a fixed strange quark mass of 100~MeV is applied
in these simulations which we want to contrast here with first results 
for a PNJL model.
A serious drawback of the bag model parametrization employed here is that 
cannot describe the mass of the recently observed 1.97~M$_\odot$
neutron star \cite{Demorest:2010bx}.
This could be cured by extending the bag model and accounting for leading order
QCD corrections and diquark condensation \cite{Ozel:2010bz,Weissenborn:2011qu}
but results in an early onset of quark matter only 
for considerably stiff nuclear EoS.

The three flavor quark matter PNJL model is based on the description in
Ref.~\cite{Blaschke:2005uj,Sandin:2007zr} with the Polyakov loop extension
according to \cite{Roessner:2006xn}. An additional isoscalar vector meson 
interaction leads to a stiffening of the quark matter equation of state and
allows to describe hybrid stars with a mass of 2~M$_\odot$ \cite{Klahn:2006iw}.
The diquark and vector meson coupling is set to $\eta_D=1.02$ and 
$\eta_V=0.25$, respectively.
Strangeness on the quark side in the PNJL model occurs at higher densities
than the onset of deconfinement, due to the dynamical quark masses involved.
The resulting phase diagram is discussed in Ref.~\cite{Blaschke:2010ka}.

\begin{figure}[h]
\subfigure[$\,\,\,\,$Quark bag model ($B^{1/4}=165$~MeV),
for
$Y_p=0.1$ (solid lines),
$Y_p=0.3$ (dash-dotted lines)
and
$Y_p=0.5$ (dotted lines)]{
\includegraphics[width=0.48\textwidth]{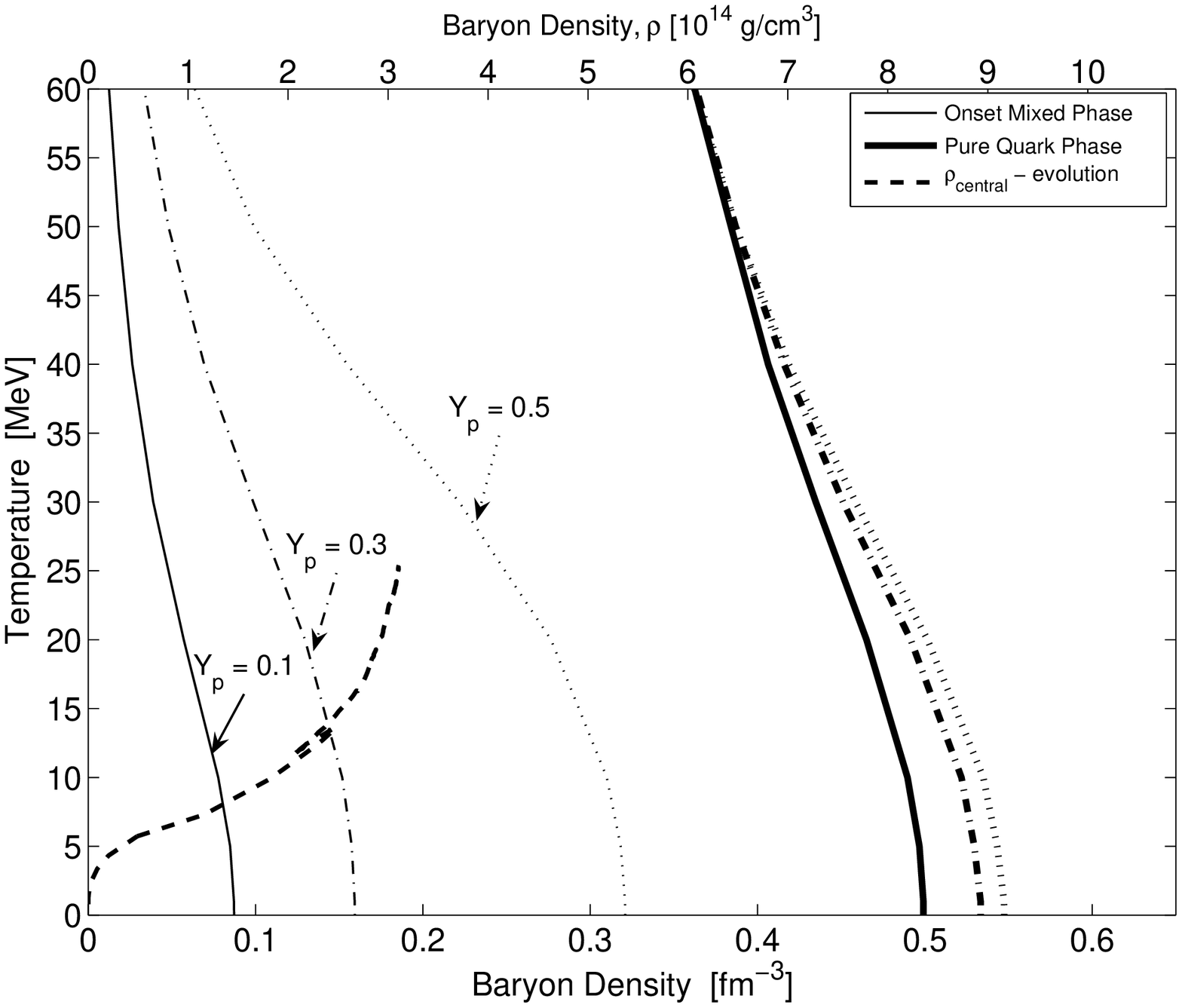}
\label{fig-phasediagram-bag}}
\hfill
\subfigure[\,\,\,\,PNJL model ($\eta_D=1.02$, $\eta_V=0.25$),  for
$Y_p=0.2$ (solid lines),
$Y_p=0.3$ (dash-dotted lines)
and
$Y_p=0.5$ (dotted lines)]{
\includegraphics[width=0.48\textwidth]{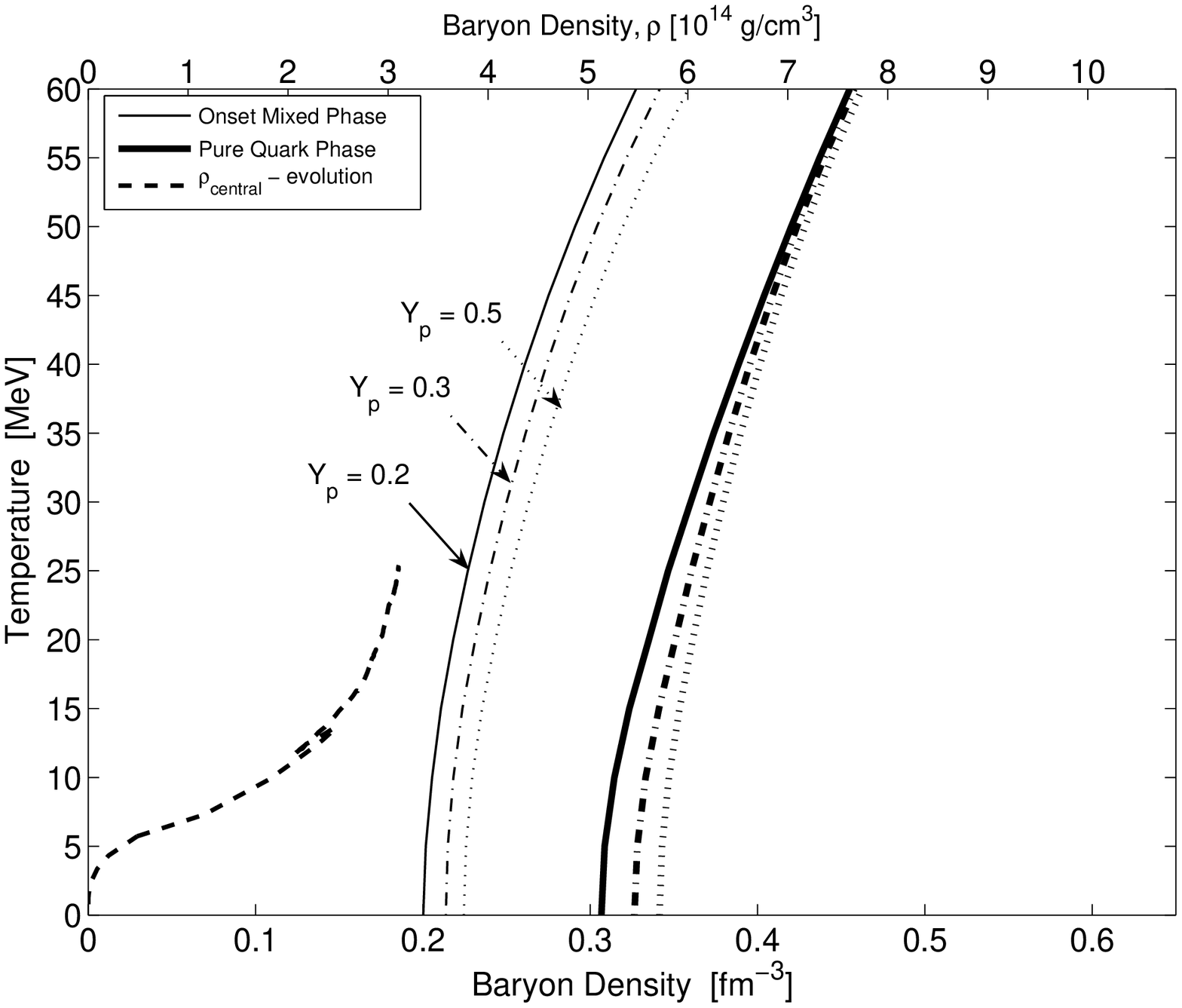}
\label{fig-phasediagram-pnjl}}
\caption[]
{QCD phase diagrams for the two hybrid EoS under discussion
(onset of quark matter: thin lines, pure quark phase: thick lines)
for selected proton-to-baryon ratios $Y_p$
that are indicated via the text arrows in the graphs.
The dashed line shows the evolution of the
central temperature and density for a representative
core collapse supernova of a 15~M$_\odot$
progenitor model \cite{Woosley:2002zz},
based on a hadronic EoS \cite{Shen:1998by}
for the first second post bounce.
Note the kink between 0.12--0.15~fm$^{-1}$, it is related to the
core bounce after which the central density and temperature decrease
slightly before they continue to increase during the later compression.
The wide stretched mixed phase region for the bag model in graph~(a)
results from the applied Gibbs construction while the in comparison
narrow region for the PNJL model in graph~(b) is a result of the Maxwell
construction.}
\end{figure}

The Figs.~\ref{fig-phasediagram-bag} and \ref{fig-phasediagram-pnjl}
compare the phase diagrams for the quark bag model and the PNJL
model for different proton-to-baryon ratios $Y_p$ relevant for
supernova matter.
There are
different critical densities for the onset of deconfinement
comparing the quark bag and PNJL models for equal $Y_p$.
E.g. for $Y_p=0.3$ and $T=0$, the critical densities are
0.159~fm$^{-1}$ for the bag model in comparison to 
0.214~fm$^{-1}$
for the PNJL model
(see the solid lines in the
Figs.~\ref{fig-phasediagram-bag} and \ref{fig-phasediagram-pnjl}).
For the bag model, the early onset of deconfinement close to
normal nuclear matter density 
illustrates the strong isospin dependency of the critical density.
For instance, at higher $Y_p=0.5$ the critical density shifts to
0.321~fm$^{-1}$ for $T=0$.
In general, small transition densities result from the application
of the Gibbs conditions for the phase transition with the corresponding
extended mixed phase and from the existence of the s-quark flavor which 
results in a larger number of degrees of freedom when compared to nuclear
matter.
For the PNJL model, the transition to quark matter,
shown in Fig.~\ref{fig-phasediagram-pnjl},
is based on the Maxwell construction.
The critical densities are generally larger and the mixed phase is very
narrow in density compared to the quark bag model.
The PNJL model also shows the typical isospin dependency.
The critical density reduces at decreasing $Y_p$, which is however
much weaker than for the quark bag model.
Both models differ also significantly in the maximum mass of neutron
stars, 1.50~M$_\odot$ for the bag model with $B^{1/4}=165$~MeV
and 1.97~M$_\odot$ for the PNJL model.

Furthermore, the temperature dependence of the critical density
is different in both quark matter descriptions.
This is of particular relevance for astrophysical applications that
explore a possible quark-hadron phase transition.
In the quark bag model, the critical density reduces continuously at
higher temperatures.
It reaches about 0.1~fm$^{-1}$ at $T=50$~MeV for $Y_p=0.3$.
For the PNJL model, the critical density rises with increasing temperature
for any $Y_p$.
This behavior is a result of the neglect of any modification of the 
Polyakov loop potential on the quark density and should be improved
as soon as reasonable constraints for such a procedure can be defined,
see \cite{Blaschke:2010vj} for a first step. 

\begin{figure}[h]
\subfigure[$\,\,\,\,$Quark bag model ($B^{1/4}=165$~MeV).]{
\includegraphics[width=0.48\textwidth]{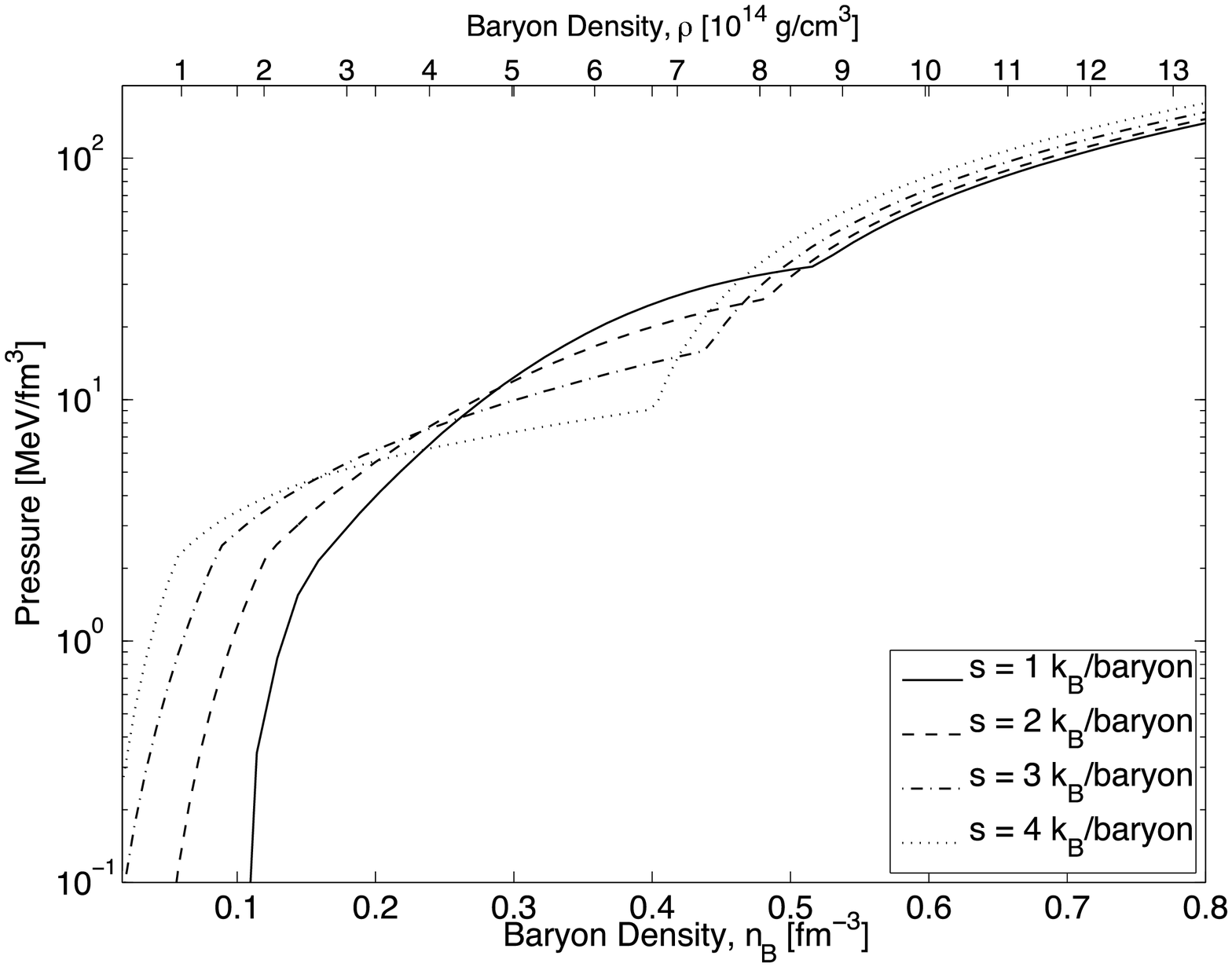}
\label{fig-pressure-bag}}
\subfigure[\,\,\,\,PNJL model ($\eta_D=1.02$, $\eta_V=0.25$).]{
\includegraphics[width=0.48\textwidth]{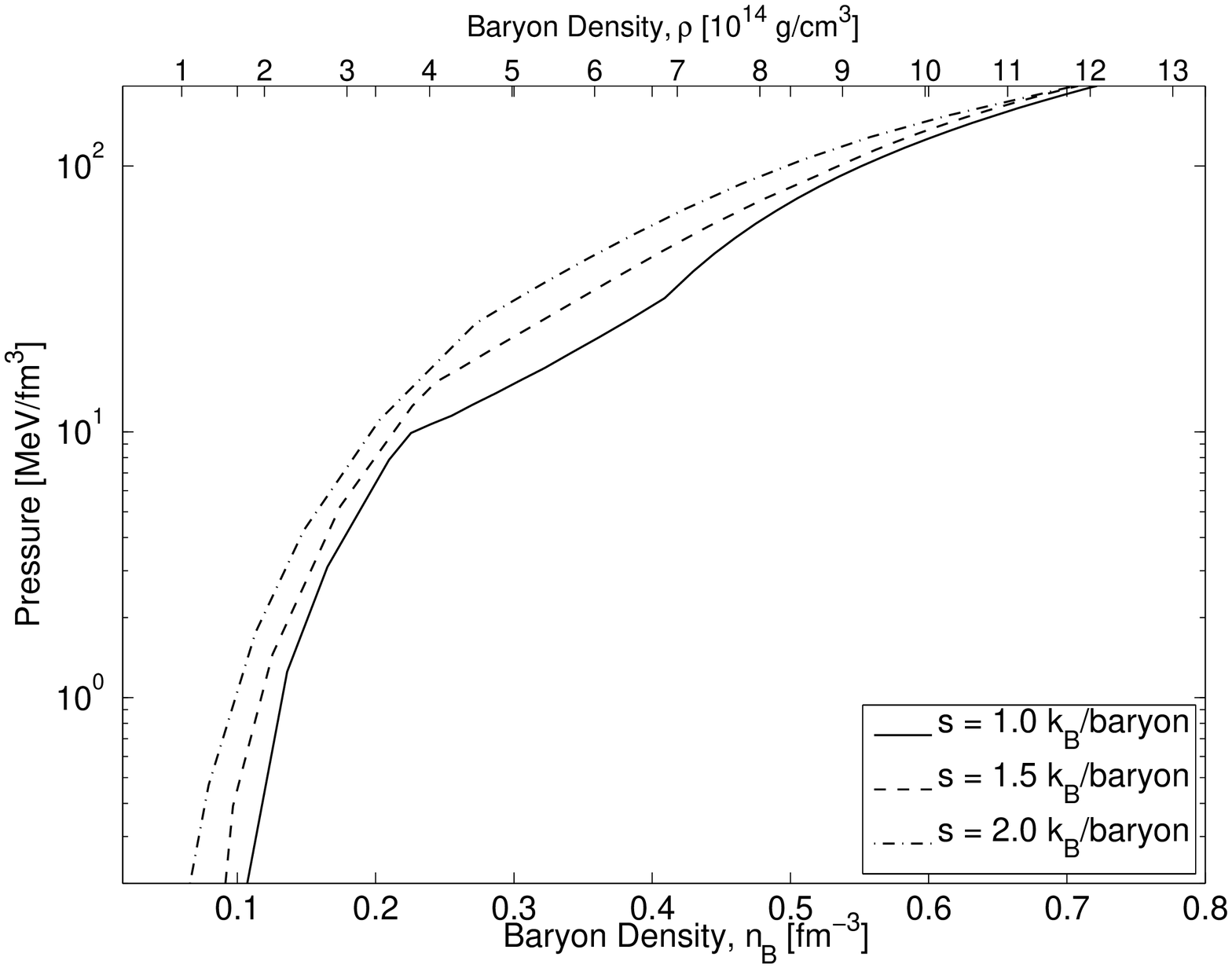}
\label{fig-pressure-pnjl}}
\caption[]
{Total pressure density curves for different entropies per baryon
and fixed $Y_p=0.3$, based on the two hybrid EoS.}
\end{figure}

The resulting EoS are illustrated in Figs.~\ref{fig-pressure-bag}
and \ref{fig-pressure-pnjl}, showing the pressure-density curves
for different entropies per baryon and fixed $Y_p=0.3$.
The largely extended mixed phase region in the phase diagram
for the bag model corresponds to a soft EoS.
This is illustrated via the pressure-density curves in
Fig.~\ref{fig-pressure-bag},
where the adiabatic index differs initially only little between
the onset of deconfinement and the pure hadronic phase.
Towards the end of the mixed phase where the pure quark phase
sets in, the adiabatic index is largely reduced for any entropy per
baryon and hence the EoS is significantly softer compared to the
hadronic case.
In the pure quark phase, the adiabatic index increases again
significantly, which makes the pure quark phase stiffer than the
mixed phase.
The sharp transition between the mixed and the pure quark phases,
gives rise to a strong effect for the hydrodynamics evolution
along isentropes.

The PNJL hybrid EoS shows a different pressure behavior
(see Fig.~\ref{fig-pressure-pnjl}) during the quark-hadron phase
transition.
The largest change of the adiabatic index is found at the onset
of deconfinement.
Close to the onset of pure quark matter, the adiabatic index
changes only little.
These two aspects are a general feature of the chosen construction
of the PNJL hybrid EoS, for which a smooth transition can be expected
in dynamical simulations.

\section{Core collapse supernova evolution in the phase diagram}

The central temperature and density evolution of the 15~M$_\odot$
core collapse supernova simulation is shown via the dashed lines in
the Figs.~\ref{fig-phasediagram-bag} and \ref{fig-phasediagram-pnjl},
for the first second post bounce, during which the explosion
is expected to take place for such massive Fe-core progenitors
\cite{Marek:2009}.
This phase of the supernova is determined by mass accretion
($\sim0.1$~M$_\odot$/s) onto the central PNS, which consequently
contracts on a timescale between 100~ms up to seconds.
During the PNS contraction, the central density and temperature
rise continuously\footnote{The highest temperatures $\sim30-60$~MeV
are not obtained at the center of the PNS but at slightly lower densities;
it corresponds to the region where the bounce shock formed initially.}.
During this evolution, the central electron fraction reduced from
$Y_p\simeq0.3$ to $Y_p\simeq0.25$.
The simulation is based on a pure hadronic description of matter
\cite{Shen:1998by}.
The current supernova models explore only hadronic EoS.
It is shown here to illustrate the possibility and the conditions
relevant in order to obtain quark matter at supernova cores.

The differences between the two hybrid EoS become clear.
Using the quark bag model, it is possible to obtain quark matter.
The early post bounce evolution of the central mass trajectories considered
for the 15~M$_\odot$ progenitor model, enter deeply into the mixed phase.
The same holds true for the non-central part of the PNS, where
up to 0.5~M$_\odot$ can reach the mixed phase within the
first 500~ms post bounce evolution
(depending on the progenitor model \cite{Fischer:2010wp}).
Within the PNJL model, the central densities obtained for this particular
15~M$_\odot$ core collapse supernova simulation are not sufficiently
high enough to enter the mixed phase.
It would require an additional rise of the central density by a factor of
two or more.
For such quark-hadron hybrid EoS quark matter will not be reached
in core collapse supernovae of massive stars up to 15~M$_\odot$
as discussed here, during the expected explosion phase of
0.5--1~seconds post bounce.

\begin{figure*}[h]
\includegraphics[width=0.99\textwidth]{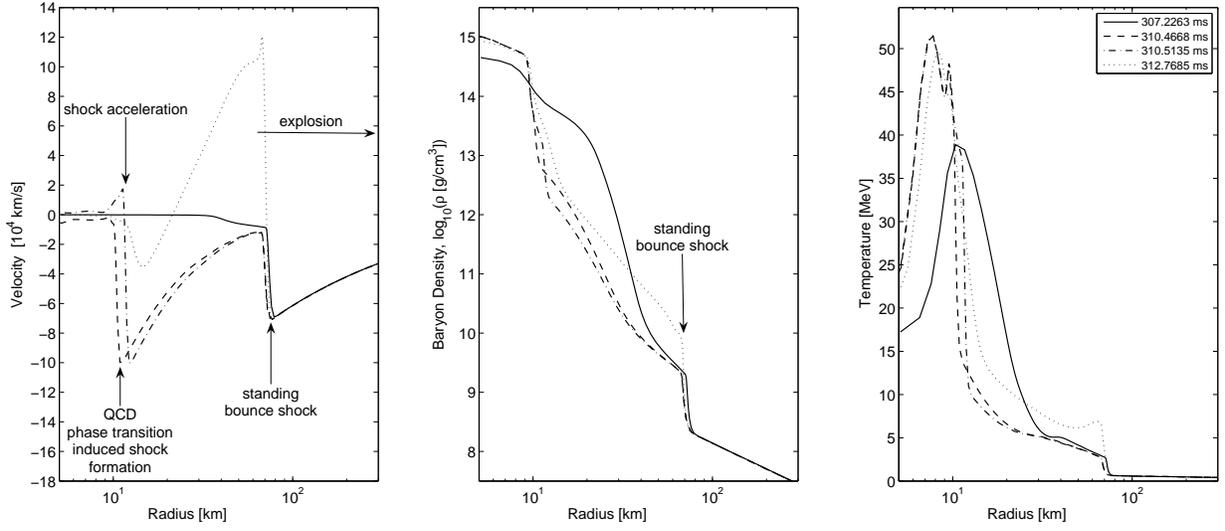}
\caption[]
{Radial profiles of velocity, density and temperature
for a 15~M$_\odot$ progenitor model at selected post bounce times
during the QCD phase transition induced PNS collapse,
using the bag model with $B^{1/4}=155$~MeV and $\alpha_S=0.3$.}
\label{fig-hydro-h15y}
\end{figure*}

In order to investigate the EoS softening effects from the presence of
quark matter in the PNS interior and possible consequences for the
dynamical evolution (and possible observations), we apply the quark
bag model hybrid EoS (using $B^{1/4}=155$~MeV, $\alpha_S=0.3$)
to a core collapse supernova evolution (again the 15~M$_\odot$ model).
The largely reduced adiabatic index at the end of the mixed phase
causes a gravitational collapse of the PNS, spatially separated into
central sub-sonic collapse and outer super-sonic collapse.
During the collapse density and temperature rise
(see Fig.~\ref{fig-hydro-h15y}),
which in turn favors pure quark matter over hadronic matter.
In the pure quark phase, where the adiabatic index is increased,
the collapse halts and a strong hydrodynamic shock front forms.
This was first observed in the context of a first order deconfinement
phase transition in ref.~\cite{Takahara:1988}, in relation to the statistically
insignificant multi-peaked neutrino signal form SN1987A \cite{Hirata:1988},
and later discussed in more detail in the refs.~\cite{Gentile:1993, Grigorian:1999}.
The shock wave appears initially as a pure accretion front, which is
illustrated in Fig.~\ref{fig-hydro-h15y} at 310.4668~ms post bounce.
The shock propagates outwards in radius towards the
PNS surface, driven by the thermal pressure of the deconfined
quarks and neutrino heating behind the shock.
The latter aspect is related to local heating rates due to electron
(anti)neutrino absorptions, which increase at the quark-hadron
phase boundary where the infalling hadronic material converts into
quark matter and density and temperature rise by several orders of
magnitude.
At the PNS surface, where the density decreases over several orders
of magnitude, the accretion front accelerates and positive velocities
are obtained
(see Fig.~\ref{fig-hydro-h15y} at 310.5135~ms post bounce).
This moment determines the onset of explosion, even in core collapse
supernova models where otherwise (i.e. without the quark-hadron phase
transition) no explosions could have been obtained.
The shock expands and continues to accelerate where matter velocities
on the order of $10^{5}$~km/s are obtained. 

\begin{figure}[h]
\includegraphics[width=0.6\textwidth]{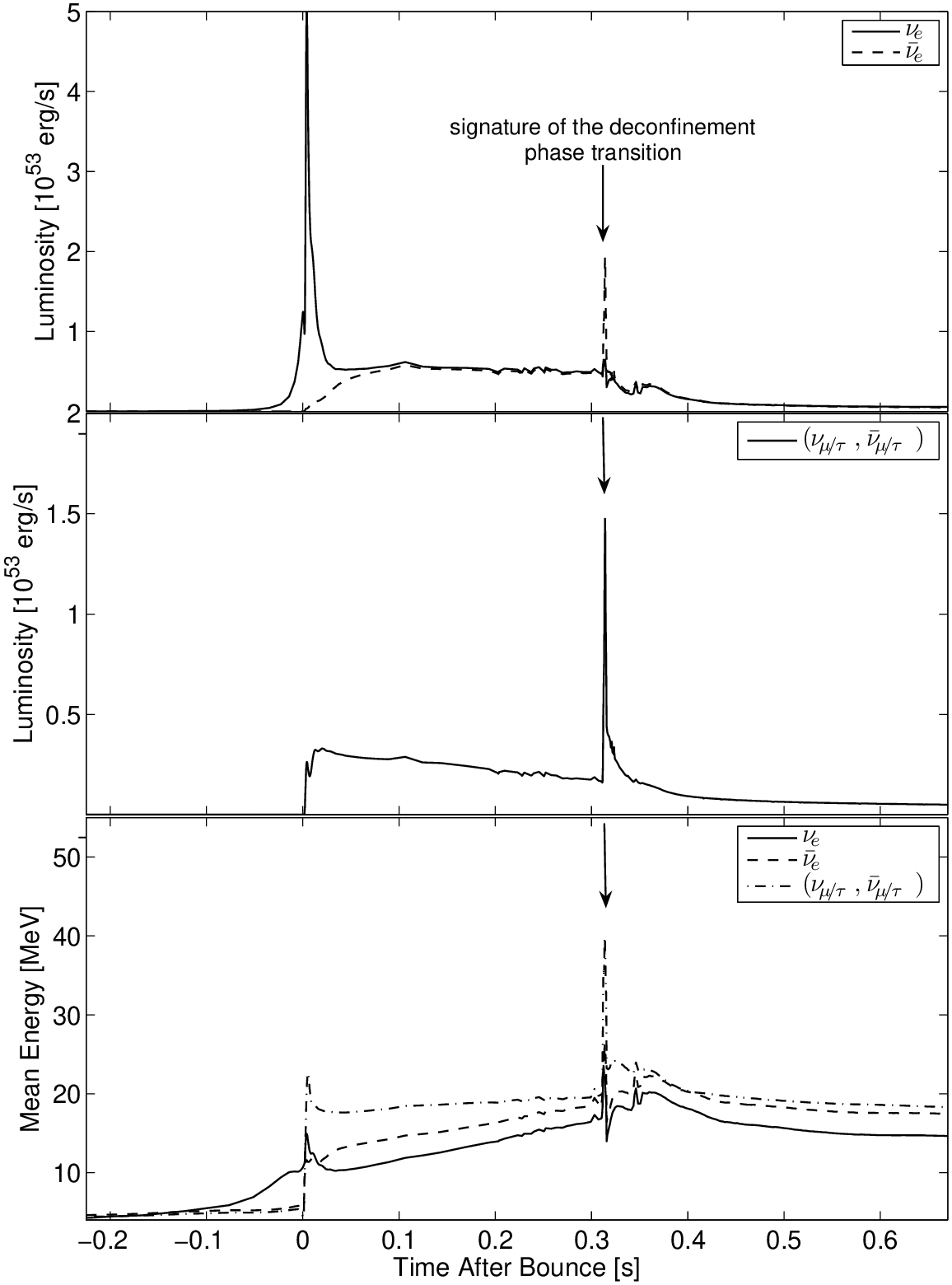}
\caption[]
{Evolution of the neutrino luminosities and mean energies
for a 15~M$_\odot$ model, including the quark hadron
phase transition using the quark bag model
($B^{1/4}=155$~MeV, $\alpha_S=0.3$)
\cite{Fischer:2010zzb, Fischer:2010wp}.
Due to the phase transition, the PNS becomes gravitational
unstable and collapses.
The critical conditions for the gravitational collapse
($T=17.67$~MeV, $\rho=5.511\times10^{14}$~g/cm$^3$, $Y_p=0.197$)
are obtained at about 308~ms after bounce for this particular model.
The second burst is released when the second shock,
that forms due to the quark hadron phase transition in the
PNS interior, crosses the neutrinospheres.
The second burst occurs in all neutrino flavors,
while the deleptonization burst at core bounce appears only
in $\nu_e$.
This signature of deconfinement is absent in models that explore
standard hadronic EoSs (e.g. from ref. \cite{Shen:1998by}).}
\label{fig-lumin}
\end{figure}

The shock propagation across the neutrinospheres releases an additional
burst of neutrinos, illustrated in Fig.~\ref{fig-lumin}.
This second burst rises in all neutrino flavors.
It differs from the reference case discussed in ref.~\cite{Fischer:2010wp},
where the second burst was dominated by $\bar{\nu}_e$ and
($\nu_{\mu/\tau}$, $\bar{\nu}_{\mu/\tau}$) due to the lower density
of the PNS envelope of the less massive progenitor chosen and
the different quark bag EoS parameters.
The rise of the burst is due to the presence of numerous electron-positron
pairs that allow for the production of $\nu_e$ and $\bar{\nu}_e$ via the
charged current reactions as well as
($\nu_{\mu/\tau}$, $\bar{\nu}_{\mu/\tau}$) via pair processes.
It differs from the deleptonization burst at core bounce, which is only
in $\nu_e$ emitted via a large number of electron captures.
This second burst, if occurring, is a strong observational indication for
the reconfiguration of the high density domain in core collapse supernova
evolution.

Coming back to the PNJL model in core collapse supernova simulations,
preliminary results of long-term non-exploding models in the progenitor
mass range between 15--25~M$_\odot$,
show a smooth transition to quark matter at the PNS interior
where only the mixed phase could be reached.
Very little mass ($\sim0.05$~M$_\odot$) is converted into the mixed phase
and the quark fraction rises only on timescales on the order of seconds,
which corresponds to the PNS contraction timescale induced via
mass accretion.
However, massive star explosions are expected to take place during
the first 0.5--1~seconds post bounce.
Hence, a direct correlation to the explosion mechanism of massive
stars as well as direct observables that relate to the explosion phase,
cannot be expected for such a hybrid EoS.
However, this hybrid EoS model has another interesting feature.
Since the PNJL model possesses a critical point, one may study whether in 
the collapse of very massive progenitors with subsequent black hole formation
the trajectory of the collapse in the QCD phase diagram sweeps the critical
point before horizon crossing \cite{Ohnishi:2011jv}.
In that case a characteristic modification of possible observables
from the deconfinement transition is expected,
at variance to trajectories which do not pass the critical point.
The realistic modeling of such processes requires further development of a
phase transition construction which does not remove the critical endpoint. 

\section{Summary}

We discussed the possibility to reach the critical conditions for the onset
of deconfinement in core collapse supernovae, comparing two different
quark-hadron hybrid EoS.
Based on the bag model for strange quark matter and choices of low
bag constants, without violating constraints from e.g. heavy-ion collision
experiments, the low critical densities as well as the strong isospin and
temperature dependencies allow for the quark-hadron 
phase transition to take place in core collapse supernova
simulations within the first 500~ms post bounce.
For a particular choice of parameters,
we illustrated the dynamical evolution.
It was determined by an adiabatic collapse due to the soft EoS in the
extended mixed phase modeled by applying Gibbs conditions,
and the formation of a strong hydrodynamic shock wave.
It triggered the explosion even in models where otherwise
explosions could not be obtained.
Furthermore, the millisecond neutrino burst released from the
shock propagation across the neutrinospheres may become
observable for a future Galactic event if quark matter occurs
\cite{Dasgupta:2009yj}.
The future observation of such a multi-peaked neutrino signal
may reveal information about the nuclear EoS at high densities and
temperatures that cannot be reached in heavy-ion collision experiments
at present. 
For the PNJL model parametrization presented here, the higher critical
density  and the narrow mixed phase due to the Maxwell construction,
together with the rising critical density for increasing temperatures,
prevent the quark-hadron phase transition to occur in core collapse
supernova simulations during the  post bounce accretion phase.
For such an EoS, a direct correlation between the phase transition to
quark matter and the explosion mechanism cannot be expected.
However, the conditions for the appearance of quark matter
for the PNJL model may allow to obtain a signal from core collapse
supernovae during the PNS deleptonization phase, on timescales
on the order of several seconds after the onset of, e.g., initially
neutrino-driven explosions.

The two classes of hybrid EoS discussed here allow for very different types
of investigations in simulations of core collapse supernovae.
Hybrid EoS based on the class of bag models for quark matter give
access to bulk properties of quark-hadron matter and allow to
model EoS which undergo a quark-hadron phase transition
during the supernova collapse accompanied by a second neutrino burst.
However, the bag model cannot capture the aspect of chiral symmetry
restoration and possible effects of the existence of a critical point.
Moreover, it remains to be shown whether the 2~M$_\odot$ mass constraint 
can be explained in agreement with the deconfinement scenario for 
this class of models.
The hybrid models based on a PNJL-type approach to quark matter would
allow to study the question of the existence of critical points in the QCD
phase diagram, e.g., during black hole formation where high temperatures
on the order of several 100~MeV are obtained. 
A systematic study of possible observables from black hole formation
using PNJL models that include a critical endpoint,
may provide the astrophysical analogue of the energy scan programs
in heavy-ion collisions.
However, the 2~M$_\odot$ mass constraint is obtained at  the price
of a too high critical density for a deconfinement transition during
collapse.
This statement is not the final word.
It is possible, e.g., via a density  dependence of the gluon thermodynamics
which is absent in the present model, to allow for a simultaneous description
of deconfinement in the supernova interior and high mass (proto)
neutron stars.  

\section*{Acknowledgment}

The authors would like to thank Hovik Grigorian for his comments and helpful discussions.  The project was funded by the Swiss National Science Foundation (SNF) under project numbers~PP00P2-124879/1, 200020-122287, and the Helmholtz Research School for Quark Matter Studies. T.F. is supported the SNF under project number~PBBSP2-133378 and by HIC for FAIR. G.M.P. is partly supported by the Sonderforschungsbereich~634, the ExtreMe Matter Institute EMMI, the Helmholtz International Center for FAIR, and the Helmholtz Association through the Nuclear Astrophysics Virtual Institute (VH-VI-417). D.B., T.K. and R.{\L}. receive support from the Polish Ministry for Science and Higher Education. D.B. acknowledges support from the Russian Fund for Basic Research under grant No. 11-02-01538-a. R.~{\L}. received support from the Bogoliubov-Infeld programme for visiting JINR at Dubna where part of his work was done. The work of G.P. is supported by the Deutsche Forschungsgemeinschaft (DFG) under Grant No.~PA~1780/2-1 and J.S.-B. is supported by the DFG through the Heidelberg Graduate School of Fundamental Physics. M.H. acknowledges support from the High Performance and High Productivity Computing (HP2C) project. S.T. is supported by the DFG cluster of excellence ''Origin and Structure of the Universe''. F.S. acknowledges support from the Belgian fund for scientific research (FNRS). I.S. is supported by the Alexander von Humboldt foundation via a Feodor Lynen fellowship and wishes to acknowledge the support of the Michigan State University High Performance Computing Center and the Institute for Cyber Enabled Research. The authors are additionally supported by CompStar, a research networking program of the European Science Foundation, and the Scopes project funded by the Swiss National Science Foundation grant.~no.~IB7320-110996/1.

\newpage

\newpage

\newpage

\newpage

\newpage

\begin{center}
FIGURE CAPTIONS
\end{center}
\begin{enumerate}
\item QCD phase diagrams for the two hybrid EoS under discussion
(onset of quark matter: thin lines, pure quark phase: thick lines)
for selected proton-to-baryon ratios $Y_p$
that are indicated via the text arrows in the graphs.
The dashed line shows the evolution of the
central temperature and density for a representative
core collapse supernova of a 15~M$_\odot$
progenitor model \cite{Woosley:2002zz},
based on a hadronic EoS \cite{Shen:1998by}
for the first second post bounce.
Note the kink between 0.12--0.15~fm$^{-1}$, it is related to the
core bounce after which the central density and temperature decrease
slightly before they continue to increase during the later compression.
The wide stretched mixed phase region for the bag model in graph~(a)
results from the applied Gibbs construction while the in comparison
narrow region for the PNJL model in graph~(b) is a result of the Maxwell
construction.
\item Total pressure density curves for different entropies per baryon
and fixed $Y_p=0.3$, based on the two hybrid EoS.
\item Radial profiles of velocity, density and temperature
for a 15~M$_\odot$ progenitor model at selected post bounce times
during the QCD phase transition induced PNS collapse,
using the bag model with $B^{1/4}=155$~MeV and $\alpha_S=0.3$.
\item Evolution of the neutrino luminosities and mean energies
for a 15~M$_\odot$ model, including the quark hadron
phase transition using the quark bag model
($B^{1/4}=155$~MeV, $\alpha_S=0.3$)
\cite{Fischer:2010zzb, Fischer:2010wp}.
Due to the phase transition, the PNS becomes gravitational
unstable and collapses.
The critical conditions for the gravitational collapse
($T=17.67$~MeV, $\rho=5.511\times10^{14}$~g/cm$^3$, $Y_p=0.197$)
are obtained at about 308~ms after bounce for this particular model.
The second burst is released when the second shock,
that forms due to the quark hadron phase transition in the
PNS interior, crosses the neutrinospheres.
The second burst occurs in all neutrino flavors,
while the deleptonization burst at core bounce appears only
in $\nu_e$.
This signature of deconfinement is absent in models that explore
standard hadronic EoSs (e.g. from ref. \cite{Shen:1998by}).
\end{enumerate}

\end{document}